\documentclass[aps,prd,twocolumn,superscriptaddress]{revtex4-1}

\usepackage{graphicx}
\usepackage{xspace}
\usepackage{amsmath}

\usepackage{color}

\usepackage[utf8]{inputenc}
\begin{document}

\title{Nuclear gluons at RHIC in a multi-observable approach}

\author{Ilkka Helenius}
\email[]{ilkka.m.helenius@jyu.fi}
\affiliation{University of Jyvaskyla, Department of Physics, P.O. Box 35, FI-40014 University of Jyvaskyla, Finland}
\affiliation{Helsinki Institute of Physics, P.O. Box 64, FI-00014 University of Helsinki, Finland}
\author{John Lajoie}
\email[]{lajoie@iastate.edu}
\affiliation{Iowa State University, Department of Physics, Ames, Iowa, 50011, USA}
\author{Joseph D. Osborn}
\email[]{jdosbo@umich.edu}
\affiliation{University of Michigan, Department of Physics, Ann Arbor, Michigan 48109, USA}
\author{Petja Paakkinen}
\email[]{petja.paakkinen@jyu.fi}
\affiliation{University of Jyvaskyla, Department of Physics, P.O. Box 35, FI-40014 University of Jyvaskyla, Finland}
\affiliation{Helsinki Institute of Physics, P.O. Box 64, FI-00014 University of Helsinki, Finland}
\author{Hannu Paukkunen}
\email[]{hannu.paukkunen@jyu.fi}
\affiliation{University of Jyvaskyla, Department of Physics, P.O. Box 35, FI-40014 University of Jyvaskyla, Finland}
\affiliation{Helsinki Institute of Physics, P.O. Box 64, FI-00014 University of Helsinki, Finland}

\date{\today}

\begin{abstract}
We explore the possibility of measuring nuclear gluon distributions at the Relativistic Heavy-Ion Collider (RHIC) with $\sqrt{s}=200 \, {\rm GeV}$ proton-nucleus collisions. In addition to measurements at central rapidity, we consider also observables at forward rapidity, consistent with proposed upgrades to the experimental capabilities of STAR and sPHENIX. The processes we consider consist of Drell-Yan dilepton, dijet, and direct photon-jet production. The Drell-Yan process is found to be an efficient probe of gluons at small momentum fractions. In order to fully utilize the potential of Drell-Yan measurements we demonstrate how the overall normalization uncertainty present in the experimental data can be fixed using other experimental observables. An asset of the RHIC collider is its flexibility to run with different ion beams, and we outline how this ability could be taken advantage of to measure the $A$ dependence of gluon distributions for which the current constraints are scarce. 

\end{abstract}

\maketitle

\section{Introduction}

Good control over the partonic structure of protons and heavier nuclei has become an indispensable ingredient in modern particle, heavy-ion, and astro-particle physics. For processes involving a large momentum transfer, $Q \gg \Lambda_{\rm QCD} \sim 200 \, {\rm MeV}$, the nucleon's relevant degrees of freedom can be described by parton distribution functions (PDFs). Despite the progress in theoretical first-principle methods \cite{Lin:2017snn}, the PDFs are still most reliably determined through a statistical analysis of a global set of experimental data. Along with the precise data from the Large Hadron Collider (LHC), the list of data types that are included in state-of-the-art PDF analyses has grown, now ranging from traditional inclusive deeply-inelastic scattering to jet, top-quark and heavy gauge-boson production \cite{Forte:2013wc,Gao:2017yyd}. At this moment, global fits of proton PDFs do not use any data from the Relativistic Heavy-Ion Collider (RHIC), and nuclear-PDF fits \cite{Paukkunen:2018kmm,Paukkunen:2017bbm} use only inclusive pion data from RHIC \cite{Adler:2006wg,Abelev:2009hx}. The advantage of the lower center-of-mass (c.m.) energies of RHIC, $\sqrt{s} = 200{\rm ~and~ }500 \, {\rm GeV}$, is that the underlying event is not as large as it is at the LHC, and thus e.g. jets can be better resolved at lower transverse momenta ($p_{\rm T}$) \cite{Li:2015gna,Adare:2015gla}. These jet measurements are compatible with next-to-leading-order (NLO) perturbative QCD calculations \cite{Ellis:1990ek,Nagy:2003tz,Adare:2015gla} down to $p_{\rm T} \sim 10 \, {\rm GeV}$ (which is the minimum $p_{\rm T}$ of the measurements), so nothing really forbids using them in PDF analysis. Similarly, low-mass Drell-Yan (DY) events can be better resolved from the decays of heavy flavour. In p+p collisions at the higher c.m. energy of $\sqrt{s} = 500 \, {\rm GeV}$, measurements of W$^\pm$ bosons also become feasible \cite{Posik:2017xok}. These measurements provide complementary constraints on the fixed-target measurements \cite{Towell:2001nh} of the $\overline{u}/\overline{d}$ ratio.

The current status of the global determination of nuclear PDFs has been recently reviewed e.g. in Refs.~\cite{Paukkunen:2018kmm,Paukkunen:2017bbm}, and the field is rapidly developing. This is mainly driven by the p+Pb measurements at the LHC, but is also motivated by theoretical advances in upgrading global analyses to the next-to-NLO (NNLO) QCD level. Currently, the most recent global NLO fits are EPPS16 \cite{Eskola:2016oht}, nCTEQ15 \cite{Kovarik:2015cma} and DSSZ12 \cite{deFlorian:2011fp}. Out of these, EPPS16 has the widest data coverage and is currently the only one to use LHC measurements. In the current NNLO-level fits \cite{Khanpour:2016pph,AbdulKhalek:2019mzd}, the experimental input is restricted to fixed-target data only. In this paper, we report our studies on the future prospects for constraining nuclear PDFs at RHIC, particularly with measurements at central and forward rapidities where forward acceptance corresponds to that proposed for the STAR \cite{Aschenauer:2016our} and sPHENIX experiments \cite{Adare:2015kwa}. Projections of forward direct photon and Drell-Yan measurements at RHIC on nuclear PDFs have been separately considered e.g. in Refs.~\cite{Arleo:2007js,Arleo:2011gc,Aschenauer:2016our}. Here, we aim for a more systematic approach by combining multiple observables into a simultaneous analysis and more carefully assessing the experimental normalization uncertainty. We will base our study mainly on the EPPS16 analysis.

\section{Experimental Data Projections}

Several processes are expected to have an impact on nuclear PDFs at RHIC c.m. energies. Here, we will focus on such double differential measurements which, to leading order, probe PDFs at fixed momentum fractions. In particular, we construct pseudodata projections for the Drell-Yan di-lepton, dijet, and direct photon-jet processes, differential in the invariant mass $M$ and rapidity $y$ of the produced pair. While the Drell-Yan process (on fixed target) has been used as a constraint for nuclear PDFs already in the pioneering EKS98 analysis \cite{Eskola:1998df}, the use of dijets \cite{Eskola:2013aya,Eskola:2019dui} has been realized only in the recent EPPS16 fit \cite{Eskola:2016oht}. Currently, there are no available photon-jet data to be included in the global analyses though the potential of the process has been discussed \cite{Stavreva:2010mw,Klasen:2017dsy}. In principle, these processes individually constrain different quark-gluon combinations of PDFs, and can also be used together to limit the effect of normalization uncertainties as will be described later. 

To generate our projections, we first impose fiducial acceptance requirements on a barrel detector with forward instrumentation. Uncertainty projections were generated for a barrel covering the full azimuth of $0<\phi<2\pi$ and pseudorapidity acceptance of $|\eta|<1$, where the detector is assumed to have full tracking as well as electromagnetic and hadronic calorimetry such that jets can be robustly measured. In conjunction with the barrel central rapidity detector, a forward spectrometer, incorporating tracking and electromagnetic and hadronic calorimetry with pseudorapidity acceptance of $1.4<\eta<4$ and full azimuthal coverage, is also considered. 

Projections were determined by taking the cross sections as predicted in the \textsc{Pythia~6} event generator~\cite{Sjostrand:2006za,Sjostrand:2007gs} and multiplying them by total integrated luminosity projections at RHIC. Assumed luminosities were 197 pb$^{-1}$ for p+p collisions and 0.33 pb$^{-1}$ for p+Au collisions, corresponding to the anticipated sPHENIX run plan for the second and third years of operation in the early 2020's. Estimates of experimental efficiencies are also applied, as described below for each process. The total expected yields were converted to per event yields by dividing by the total p+p cross section times the expected luminosity. Thus, the ratio of the p+Au to p+p yields is always unity and the statistical uncertainties of the ratio are indicative of the actual statistical uncertainties on a measurement of $R_{\mathrm{p}A}$, where $R_{\mathrm{p}A}$ is defined as 
\begin{equation}
    R_{\mathrm{p}A} \equiv \frac{1}{A} \frac{\mathrm{d}\sigma_{\mathrm{p}A}/\mathrm{d}y\mathrm{d}M^2}{\mathrm{d}\sigma_{\mathrm{pp}}/\mathrm{d}y\mathrm{d}M^2}
\end{equation}

Since the detector has both central and forward instrumentation, there are a number of rapidity regions that each observable can probe. Ideally measurements should be made in each region, as different values of $x$ will be probed in both the proton and the nucleus when measuring at central and/or forward rapidities. Thus, we consider several combinations of observables, for which a summary table is shown in Tab.~\ref{tab:data_summary}. 

Drell-Yan data are generated in both the central barrel and the forward arm independently from one another. In each case, the Drell-Yan dilepton pair is measured in the invariant mass range of $4.5<M_{\ell^+\ell^-}<9$ GeV/$c^2$. Experimental efficiencies for the reconstruction of Drell-Yan pairs were estimated from a full {\textsc{Geant4}}~\cite{Agostinelli:2002hh} simulation of the sPHENIX detector (including forward instrumentation). Tight cuts on the simulated data were used to reduce backgrounds so that the simulated measurement is dominated by Drell-Yan pairs and backgrounds from decays, conversions, etc. were minimal. Overall pair efficiencies vary between 10-15\% over the invariant mass range considered for both central and forward measurements. The dijet data are considered in both regions to probe a variety of $x$ values. Jets were determined from final-state \textsc{Pythia} particles with the anti-$k_{\rm T}$ algorithm with $R=0.4$~\cite{Cacciari:2011ma}. Two dijet pseudodata samples are constructed, one in which both jets are measured in the central barrel and another where one jet is measured at central rapidity and the other is measured at forward rapidity. Since the $p_T$ reach of jets becomes smaller at forward pseudorapidities, the leading jet at central rapidity is required to have $p_{\rm T}>12$ GeV/$c$ and the subleading jet at forward rapidity is required to have $p_{\rm T}>8$ GeV/$c$. Experimental efficiencies for the reconstruction of dijet pairs were estimated in a similar fashion as the Drell-Yan data. Jets were reconstructed with both hadronic and electromagnetic calorimeter deposits and efficiencies were found to be approximately 80\%, where these efficiencies become smaller towards the edge of the detector acceptance when some fraction of the jet cone lies outside of the detector. At forward pseudorapidity, the efficiencies were generally smaller, varying between 40-70\% depending on the pseudorapidity of the jet. The direct photon-jet channel is expected to have high impact on the gluon nuclear PDF at small $x$ when the process is measured in the forward direction. However, the photon-jet channel is difficult to measure at forward pseudorapidities due to large backgrounds from $\pi^0\rightarrow\gamma\gamma$ decays. Thus, we only generate photon-jet pseudodata where both are measured in the central barrel, where previous direct photon measurements at RHIC have been made and where future RHIC experimental upgrades are expected to be able to measure this process. We note that if new forward instrumentation at RHIC were available to separate direct photons from backgrounds at forward pseudorapidities this would add a powerful additional observable to constrain the nuclear gluon PDF at low $x$, complementary to Drell-Yan \cite{Aschenauer:2016our}. The photon-jet cross sections were generated with $p_{\rm T}^\gamma>10$ GeV/$c$ and $p_{\rm T}^{\rm jet}>$ 8 GeV/$c$. Photon-jet reconstruction efficiencies were evaluated similarly to the dijet and Drell-Yan data, where the efficiency was found to be approximately 70\% integrated across the central rapidity of the barrel detector.

 \begin{table}[tbh]
 \caption{\label{tab:data_summary} A summary table showing the different combinations of pseudorapidity measurements for each channel generated in this study.}
 \begin{ruledtabular}
 \begin{tabular}{ccc}
 Central-central & Forward-central & Forward-forward \\
 \hline
 Drell-Yan & & Drell-Yan \\
 Dijets & Dijets & \\
 Photon-jet & & \\
 \end{tabular}
 \end{ruledtabular}
 \end{table}

\subsection{Generation of pseudodata}

From the \textsc{Pythia} simulations for p+p and p+Au collisions we keep the relative statistical error, but construct the pseudodata points for the expected nuclear modification $R_{\mathrm{p}A}$ as
\begin{equation}
R_{\mathrm{p}A} = R_{\mathrm{p}A}^{\mathrm{EPPS16}} \times \left[1 + r \delta^{\rm uncorr.}\right] \,, \label{eq:psudodatagen}
\end{equation}
where $\delta^{\rm uncorr.}$ signifies the total uncorrelated data uncertainty and $r$ is a Gaussian random variable. To obtain $\delta^{\rm uncorr.}$, we add in quadrature the statistical uncertainty in the anticipated yield in p+p and p+Au collisions. A 4\% normalization uncertainty is assumed to account for the model dependence in determining $\langle N_{\rm coll}\rangle$ used in determining the $R_{\mathrm{p}A}$ ratio.  The overall scale of this uncertainty is unimportant, however, assuming it is common to all measurements, as we will detail later. In addition, for dijet (photon-jet) measurement, another 5\% (4\%) uncorrelated bin-to-bin systematic uncertainty is added, corresponding to the residual experimental systematic error that does not cancel in the ratio. For the Drell-Yan case the statistical uncertainty dominates and no additional systematic uncertainty is added. A systematic uncertainty of the order of 5\% is clearly smaller than what one can expect to be present in measurements for the absolute cross sections. However, if the p+p and p+Au runs are made soon one after the other (so that the detector configuration and calibration remains unaltered), much of the systematic uncertainty can be expected to cancel. We note that recent dijet measurements by the CMS collaboration \cite{Sirunyan:2018qel} quote a systematic uncertainty even less than 5\%.

The central values for $R_{\mathrm{p}A}$ in Eq.~(\ref{eq:psudodatagen}) were obtained by NLO-level calculations using the CT14NLO \cite{Dulat:2015mca} free-proton PDFs and EPPS16 \cite{Eskola:2016oht} nuclear modifications. For dijets we used \textsc{meks} (v1.0) \cite{Gao:2012he}, with the anti-$k_{\rm T}$ algorithm, taking a jet cone $R=0.4$, and fixing the QCD scales to the average of the two highest-$p_{\rm T}$ jets. The leading jet was required to have $p_{\rm T}>12$ GeV/$c$ and the subleading jet $p_{\rm T}>8$ GeV/$c$. These unequal cuts are necessary to avoid sensitivity to soft-gluon resummation. For photon-jet we used \textsc{jetphox} (v1.3.1) \cite{Catani:2002ny,Belghobsi:2009hx} where the jet was defined by a $k_{\mathrm{T}}$ algorithm with $R=0.4$ and the QCD scales were fixed to $p_{\mathrm{T}}$ of the photon. No isolation criteria for the photons were imposed.
The NLO Drell-Yan cross sections are standard, and were calculated with a private code based on Ref.~\cite{Paukkunen:2009ks}, fixing the QCD scales to the invariant mass of the dilepton pair. Besides the photon-jet process for which there are no data in the EPPS16 analysis, the scale choices are the same as made in the EPPS16 fit.

\section{Impact on EPPS16}

\subsection{The Hessian reweighting technique in a nutshell}
\label{sec:Hesse}

We estimate the impact of the projected data on the EPPS16 nuclear PDFs by the PDF reweighting (also called PDF profiling) method \cite{Paukkunen:2013grz,Paukkunen:2014zia,Camarda:2015zba, Schmidt:2018hvu,Eskola:2019dui}. In this method, one studies the function
\begin{equation}
\chi^2(\vec z) = \sum_i \left(a_iz_i^2 + b_iz_i^3 \right) + \chi^2_{\rm new \ data}(\vec z) \,, \label{eq:overallchi2}
\end{equation}
where the first term describes the behaviour of the original global $\chi^2$ in the EPPS16 analysis, and the second term is the contribution of the new data to the overall $\chi^2$ budget. The central fit of EPPS16 corresponds to $\vec z = \vec 0$, and the error sets $S_i^\pm$ are defined in the $z$ space by
\begin{align}
S_1^\pm & = (\delta_1^\pm, 0, \ldots, 0) \nonumber \\
S_2^\pm & = (0,\delta_2^\pm, \ldots, 0) \nonumber \\
        & \vdots \label{eq:errorsets} \\
S_N^\pm & = (0,0, \ldots, \delta_N^\pm) \,, \nonumber
\end{align}
and they are known to increase the original $\chi^2$ function by $T = 52$ units. The values for $\delta_i^\pm$ are given in the EPPS16 paper \cite{Eskola:2016oht}, from which $a_i$ and $b_i$ coefficients in Eq.~(\ref{eq:overallchi2}) can be solved. The contribution from the ``new'' data is defined as
\begin{equation}
\chi^2_{\rm new \ data}(\vec z) = \sum_i \left[\frac{D_i-f_NT_i(\vec z)}{E_i} \right]^2 + \left[\frac{f_N-1}{E^{\rm norm.}}\right]^2 \,, \label{eq:chinewdata}
\end{equation}
where $D_i$ and $E_i$ denote the $i$th data point and it's error. The overall normalization uncertainty is marked by $E^{\rm norm.}$. The theoretical prediction $T_i(\vec z)$ we write as
\begin{equation}
T_i(\vec z) = T_i(\vec z = \vec 0) + \sum_k \left[ \beta_{ik} z_k + \gamma_{ik} z_k^2 \right] \,,
\end{equation}
where the coefficient for $\beta_{ik}$ and $\gamma_{ik}$ can be obtained by computing the predictions for $T_i$ with all the PDF error sets. As a result, the total $\chi^2$ in Eq.~(\ref{eq:overallchi2}) becomes an analytic function of $\vec z$ which we numerically minimize with respect to $\vec z$ and $f_N$. We note that to avoid D'Agostini bias \cite{DAgostini:1993arp}, the normalization factor in Eq.~(\ref{eq:chinewdata}) $f_N$ multiplies the theoretical prediction $T_i$, and not the data value $D_i$. Since the pseudodata are based on EPPS16, the new minimum is, by construction, always very close to $\vec z = \vec 0$. After finding the parameters $\vec z_{\rm min}$ that correspond to the minimum of Eq.~(\ref{eq:overallchi2}), we expand 
\begin{equation}
\Delta\chi^2(\vec z) \equiv \chi^2(\vec z) - \chi^2(\vec z_{\rm min})  \approx  (\vec z - \vec z_{\rm min})^{\rm T} H (\vec z - \vec z_{\rm min}) \,, \label{eq:overallchi2after}
\end{equation}
where $H$ is the second-derivative (Hessian) matrix. By diagonalizing the matrix $H$ this becomes
\begin{equation}
\Delta\chi^2(\vec z) = \Delta\chi^2(\vec v)  \approx (\vec v)^2 \,, 
\end{equation}
where $\vec v = P (\vec z - \vec z_{\rm min})$, where $P$ is the orthogonal matrix that diagonalizes $H$, i.e. $P^{T} H P = 1$. The new error sets are then defined as in Eq.~(\ref{eq:errorsets}) assuming that the original tolerance is not altered, i.e. that each new error set $\hat S_i^\pm$ still correspond to $\Delta\chi^2(\hat S_i^\pm) = 52$.

\subsection{Correlating the overall normalization}

The normalization uncertainty in Eq.~(\ref{eq:chinewdata}) we discuss here is that of the luminosity determination of the minimum-bias data sample. At the LHC, the p+$A$ luminosities are determined by Van der Meer scans \cite{Grafstrom:2015foa}. Alternatively, the measured per-event yields $\mathrm{d}N^{\mathrm{p}A}/N_{\rm events}$ are converted to cross sections $\mathrm{d}\sigma^{{\rm p}A}$ by
\begin{equation}
\mathrm{d}\sigma^{{\rm p}A} = \frac{\sigma^{\rm inelastic}_{\rm pn}}{\langle N_{\rm coll} \rangle} \frac{\mathrm{d}N^{{\rm p}A}}{N_{\rm events}} \,, \label{eq:yieldstoxsecs}
\end{equation}
where the average number of binary nucleon-nucleon collisions $\langle N_{\rm coll} \rangle$ is estimated from a Glauber-type model \cite{Miller:2007ri}. This leads to a model-dependent normalization uncertainty which is difficult to determine. Furthermore, the inelastic proton-nucleon cross section $\sigma^{\rm inelastic}_{\rm pn}$ appearing in Eq.~(\ref{eq:yieldstoxsecs}) is very sensitive to the physics at low scales $Q^2 \sim \Lambda_{\rm QCD}^2$, and it is presumably lower than the values measured in proton-proton collisions due to shadowing/saturation effects. The overall normalization is thus problematic in this approach. However, the normalization issue can be overcome by simultaneously measuring several observables from the same minimum-bias data sample. The reason is that there is only one single normalization uncertainty and in Eq.~(\ref{eq:chinewdata}) the index $i$ runs through all data points, not just those belonging to a one single observable. Including data that probe PDFs in a relatively better constrained region thus serves to ``calibrate'' the overall normalization.

\begin{figure}[htb!]
    \centering
    \includegraphics[width=1.00\linewidth]{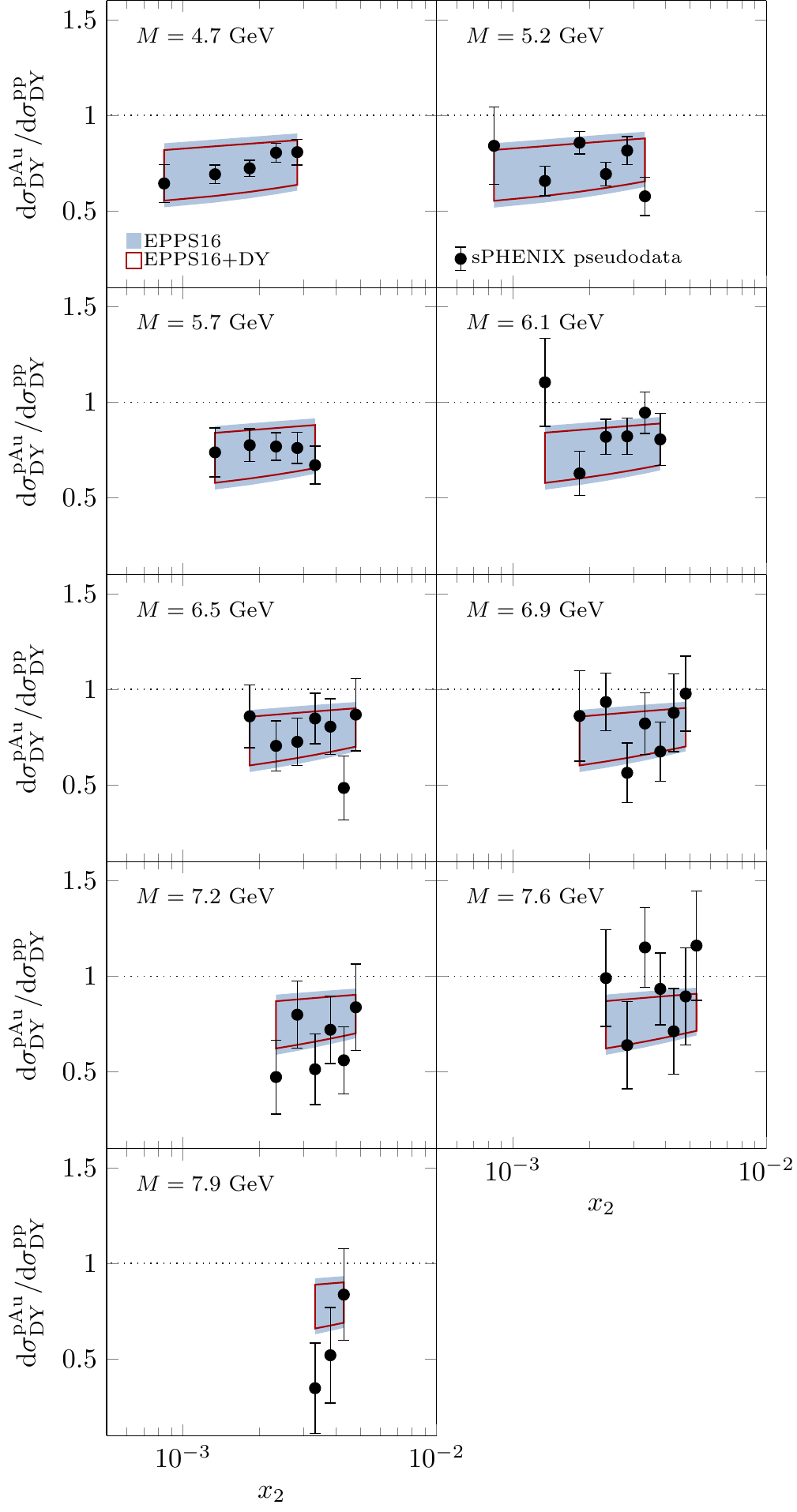}
    \caption{Effect of PDF reweighting when only the forward Drell-Yan data (shown in the plot) are used in the analysis. The light-blue bands denote the original EPPS16 uncertainties, and the red lines indicate the new upper and lower uncertainty limit after reweighting.}
    \label{fig:onlyfwdDY}
\end{figure}

To demonstrate how this works we have performed a PDF-profiling analysis first using only the forward Drell-Yan data, and then supplementing these data with the central-barrel dijet data. When only the Drell-Yan data are used, the constraints appear very weak. This is shown in Fig.~\ref{fig:onlyfwdDY} where the original EPPS16 uncertainties on the predictions are barely affected by the reweighting. The reason for the inability of these data to provide constraints is that the nuclear modification is predicted to be rather flat at small $x$ and the variations in PDFs around the original central value can be compensated by suitably tuning the normalization $f_N$. The flatness of the predicted Drell-Yan nuclear modification originates, to some extent, from the fit functions used in the EPPS16 analysis, but also from the scale evolution of PDFs which tends to flatten out the nuclear modifications in sea-quarks. Here, we took the normalization uncertainty to be 4\%, but if a larger number would have been used (e.g. 10\%) even fewer constraints would have been obtained. 
\begin{figure}[htb!]
    \centering
    \includegraphics[width=1.00\linewidth]{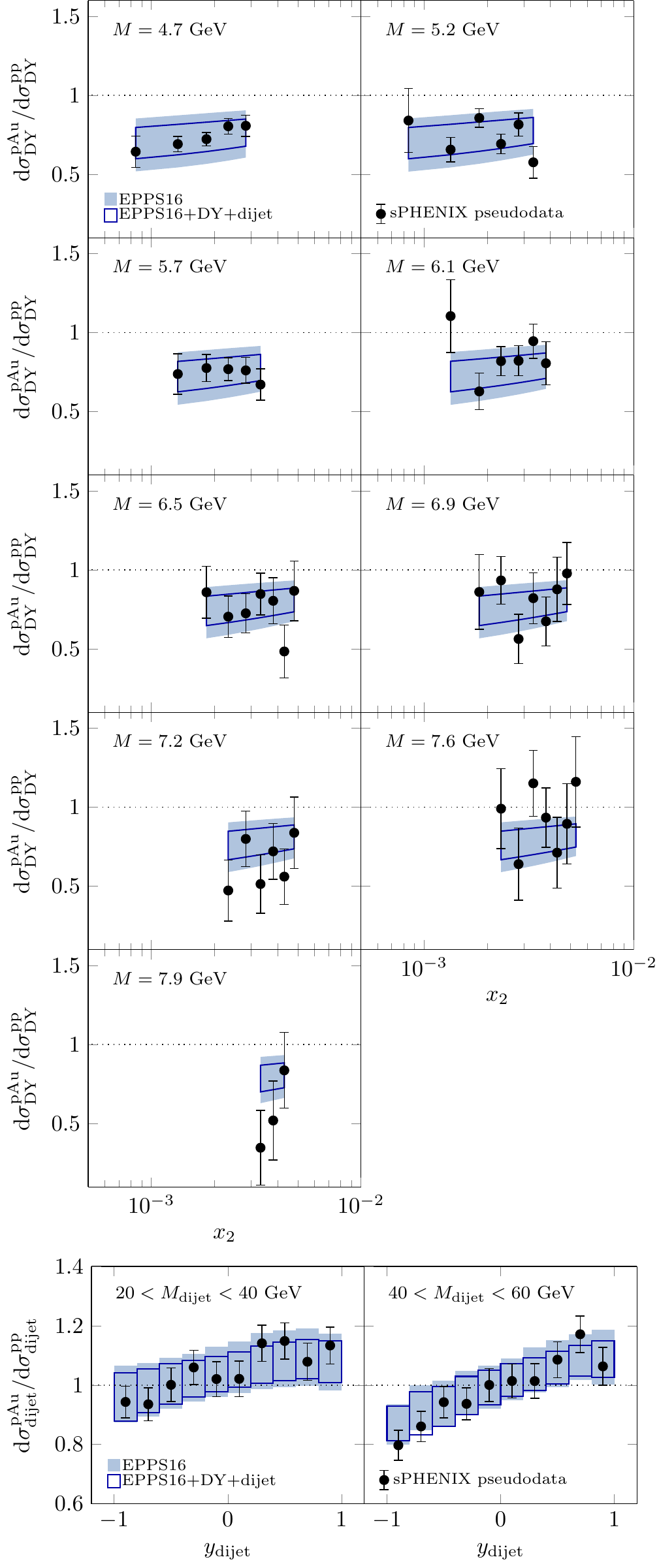}
    \caption{Effect of PDF profiling using both the forward Drell-Yan (upper panels) and central-barrel dijet data (lower panels) with common normalization. The dark-blue lines indicate the new upper and lower uncertainty limit after the PDF profiling.}
    \label{fig:fwdDYcentjet}
\end{figure}
\begin{figure}[htb!]
    \centering
    \includegraphics[width=0.99\linewidth]{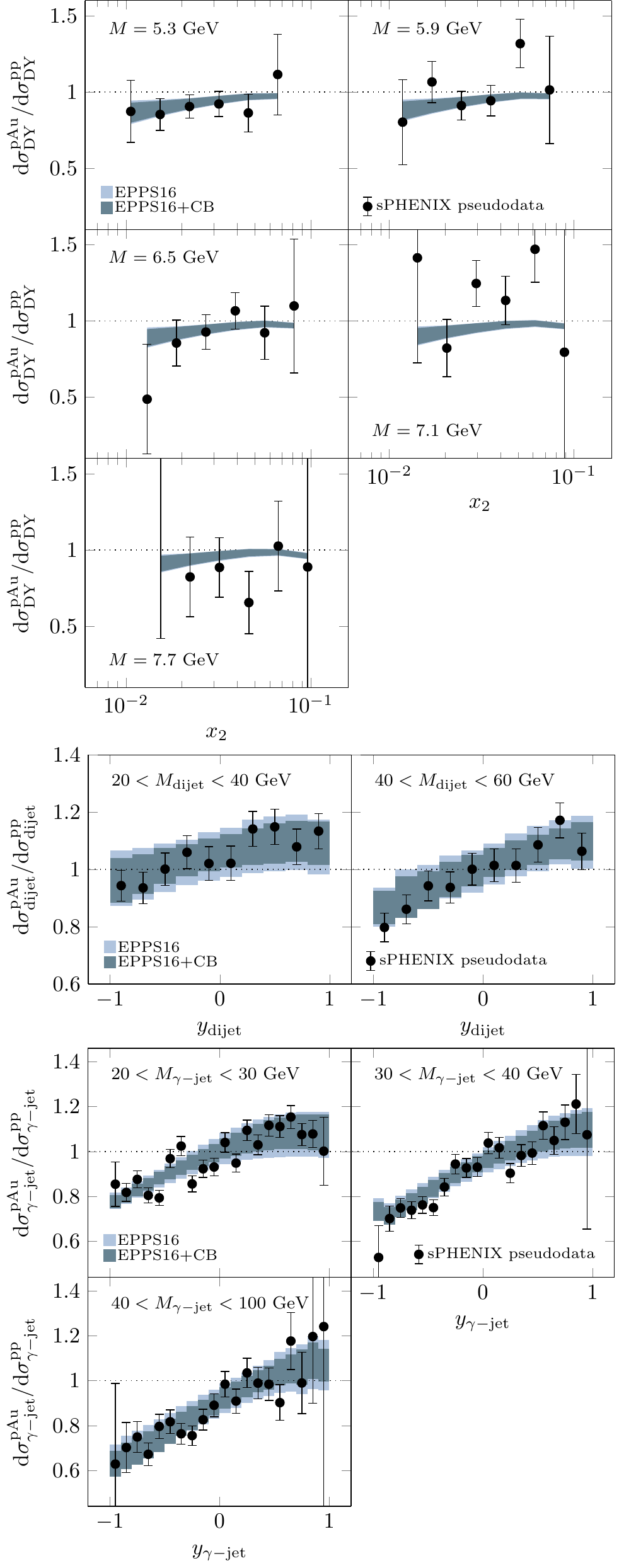}    \caption{Effect of PDF profiling using simultaneously the central-barrel Drell-Yan (upmost panels), dijet (middle panels), and photon-jet (bottom panels) pseudodata.}
    \label{fig:cbarrelfit}
\end{figure}
\begin{figure*}[htb!]
    \centering
    \includegraphics[width=1.00\linewidth]{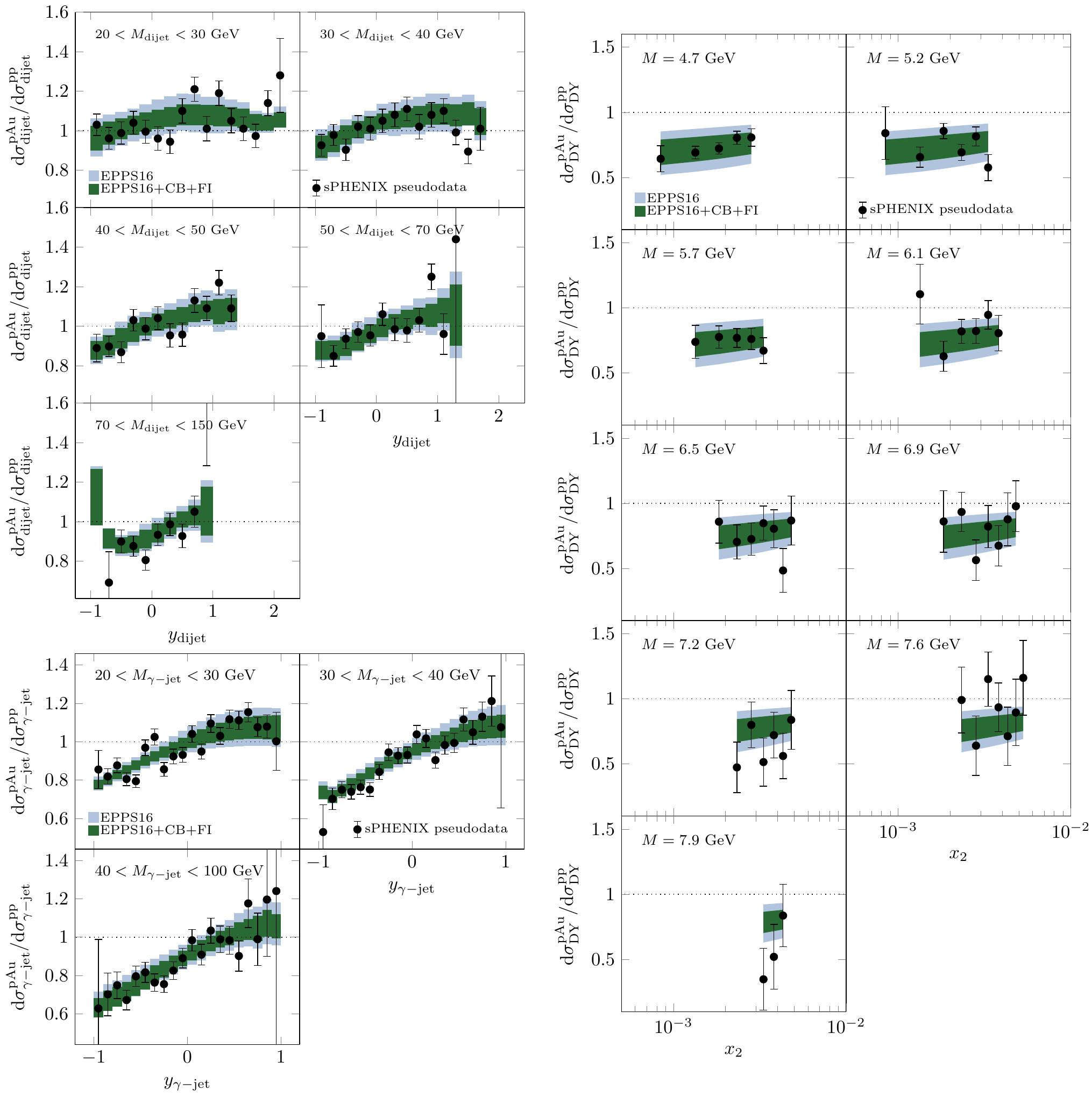}    \caption{Profiling analysis using simultanously the full dijet (upper left panels), photon-jet (lower left panels), and Drell-Yan (rightmost panels) pseudodata. Central-barrel Drell-Yan pseudodata are omitted from the figure.}
    \label{fig:fullfit}
\end{figure*}
\begin{figure}[htb!]
    \centering
    \includegraphics[width=1.00\linewidth]{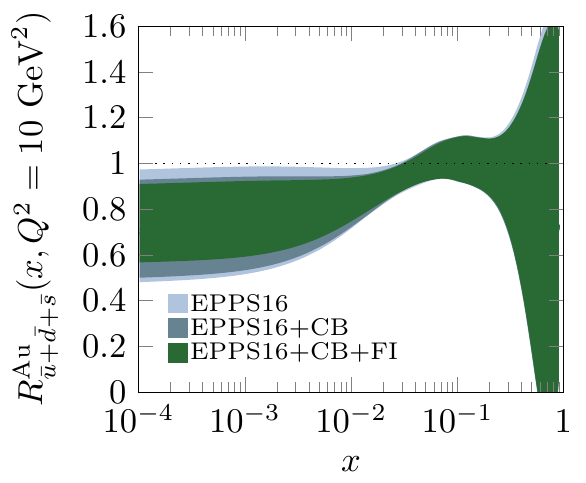} \\
    \includegraphics[width=1.00\linewidth]{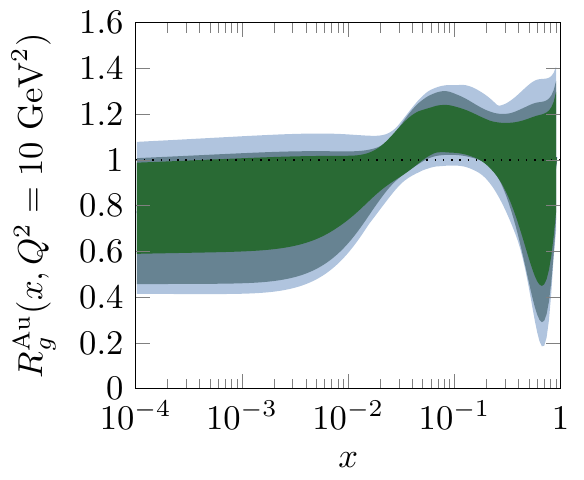}
    \caption{Effects of PDF profiling for EPPS16. The light-blue bands ("EPPS16") are the original EPPS16 errors and the darker bands ("EPPS16+CB") those after profiling with the central-barrel pseudodata. The results of adding also the forward-instrumentation data are shown in green bands ("EPPS16+CB+FI").}
    \label{fig:fit_nPDF}
\end{figure}

The situation changes when the central-barrel dijet projections are also included. These data probe the nuclear PDFs at much higher $x$ than the Drell-Yan data and carry significant sensitivity also to the rather-well constrained sum of valence quarks $u^A_{\rm valence} + d^A_{\rm valence}$. The nuclear modifications for the dijets are expected to exhibit some excess (antishadowing) around $y_{\rm dijet} \sim 1$ which turns into a suppression (EMC effect) for $y_{\rm dijet} \sim -1$. Such a pattern can not be mimicked by the overall normalization and leaves thus less room for $f_N$ variation. Since the normalization is now common for the dijet and Drell-Yan data, the Drell-Yan data have much larger impact. This is shown in Fig.~\ref{fig:fwdDYcentjet}, which should be compared to Fig.~\ref{fig:onlyfwdDY}. While the uncertainties for the central-barrel dijet data are only slightly reduced from their original EPPS16 values, the inclusion of these data is crucial in fixing the overall normalization. We have also observed that our results do not significantly depend on the exact value we pick for the normalization uncertainty.

\subsection{Simultaneous analysis of Drell-Yan, dijet and photon-jet pseudodata}

Following the observation made in the previous subsection, our strategy is to simultaneously analyze several observables that share the common normalization uncertainty. To separate the effect of forward-arm measurements, we first present the results using only the central-barrel data, and then include the data simulated with the forward-arm acceptance. 

In Fig.~\ref{fig:cbarrelfit} we summarize the Drell-Yan, dijet and photon-jet pseudodata within the central-barrel acceptance $-1 < \eta < 1$. The light-blue bands (``EPPS16'') show the original EPPS16 predictions, and the darker bands (``EPPS16+CB'') the error bands obtained after the reweighting analysis. We observe a modest improvement in the uncertainty bands for dijet and $\gamma$-jet cases. The precision of the Drell-Yan measurements is not expected to be high enough to set constraints as the  sea quarks at $10^{-2}<x<10^{-1}$ are already rather well constrained by the fixed-target DIS data. In Fig.~\ref{fig:fullfit}, in turn, we show the combined pseudodata within the full central-barrel and forward-instrumentation acceptance \footnote{We note that it would still be possible to increase the acceptance towards large nuclear $x$ by swapping the beam directions (p-$A$ $\leftrightarrow$ $A$-p) as has been done at the LHC.}. The light blue bands are again the original EPPS16 predictions, and the green bands (``EPPS16+CB+FI'') are the uncertainties obtained in the reweighting exercise. The reduction in the PDF uncertainties is now more significant than in the central-barrel-only case shown in Fig.~\ref{fig:cbarrelfit}.

The impact of both ``EPPS16+CB'' and ``EPPS16+CB+FI'' analyses on EPPS16 is shown in Fig.~\ref{fig:fit_nPDF} where we plot the average sea-quark modification for Au,
\begin{align}
    & R_{\overline{u}+\overline{d}+\overline{s}}^{{\rm Au}}(x,Q^2) \equiv \\ & \frac{f_{\overline{u}}^{\rm p/{\rm Au}}(x,Q^2)+f_{\overline{d}}^{\rm p/{\rm Au}}(x,Q^2)+f_{\overline{s}}^{\rm p/{\rm Au}}(x,Q^2)}{f_{\overline{u}}^{\rm p}(x,Q^2)+f_{\overline{d}}^{\rm p}(x,Q^2)+f_{\overline{s}}^{\rm p}(x,Q^2)} \,, \nonumber
\end{align}
together with the gluon nuclear modification,
\begin{align}
    & R_g^{{\rm Au}}(x,Q^2) \equiv \frac{f_g^{\rm p/{\rm Au}}(x,Q^2)}{f_g^{\rm p}(x,Q^2)} \,. 
\end{align}
Here, $f_i^{\mathrm{p}/A}(x,Q^2)$ denotes the parton density in a bound proton and $f_i^{\rm p}(x,Q^2)$ is the free-proton PDF. We omit here the valence quarks as we found no effects there. The improvement we find in $R_{\overline{u}+\overline{d}+\overline{s}}^{A}$ is rather weak in both cases. In the central-barrel analysis, there is a modest improvement in the gluons across all values of $x$, though the small-$x$ improvement is merely a consequence of better constrained anti-shadowing regime which is transmitted to small $x$ via momentum conservation and correlations in the EPPS16 fit function. The improvement for $R_g^{A}(x,Q^2)$ in the full analysis is clearly larger. Thanks to the wider acceptance, the full pseudodata sample is able to provide better direct constraints also at lower $x$. In particular, the gluon distribution gets significantly better constrained, the level of improvement being of the order of 50\% in places. 

\begin{figure}[htb!]
    \centering
    \includegraphics[width=1.00\linewidth]{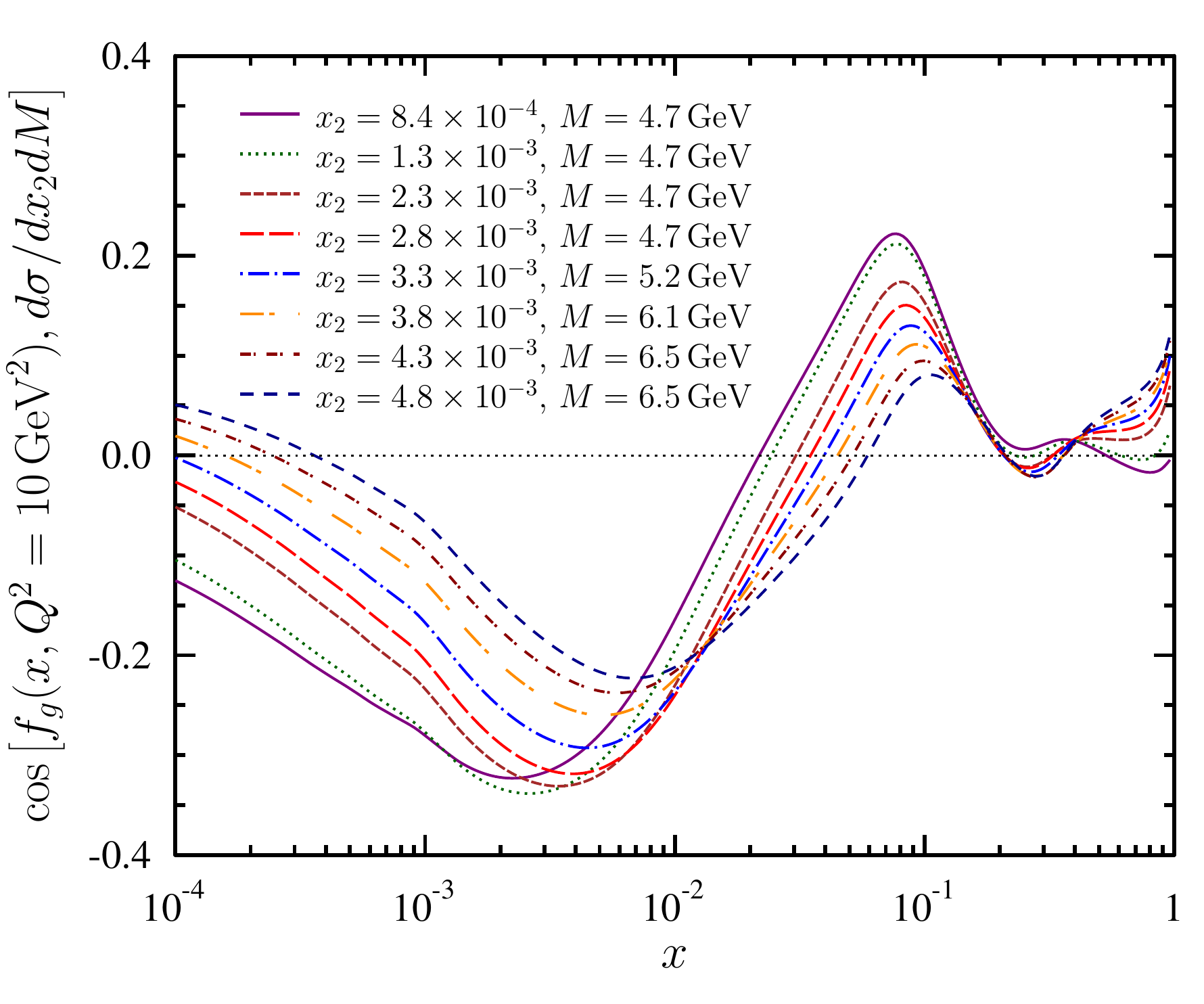} \\
    \caption{Correlation cosine between the gluon PDF at $Q^2=10\,{\rm GeV}^2$ and small-$x_2$ Drell-Yan cross sections. }
    \label{fig:correlationcosine}
\end{figure}

Here we have found that the most decisive factor for constraining the small-$x$ gluons is the forward-arm Drell-Yan data sample. At leading order, the Drell-Yan production occurs only via $q\overline{q}$ annihilation, but at small $x$ the behavior of sea quarks is still strongly driven by the gluons. At NLO and beyond there is, in addition, a direct gluon contribution from the quark-gluon scattering. To further illustrate the sensitivity of the Drell-Yan process to the gluon PDF, Fig.~\ref{fig:correlationcosine} shows examples of the correlation cosine \cite{Pumplin:2001ct} between the gluon PDF and the Drell-Yan cross sections at fixed forward kinematics. Using the notation of Sect.~\ref{sec:Hesse}, the correlation cosine of two quantities $X$ and $Y$ is defined as
\begin{align}
\cos\left(X,Y\right) & \equiv \frac{\sum_k \Delta X_k \Delta Y_k}{\left(\sum_i \Delta X_{i}^2 \right) \left(\sum_k\Delta Y_k^2\right)} \,, \label{eq:correl} \\[5pt]
\Delta X_k & \equiv X(S_k^+)-X(S_k^-) \,, \\[5pt]
\Delta Y_k & \equiv Y(S_k^+)-Y(S_k^-) \,.
\end{align}
We take $X=f_g(x,Q^2)$ and $Y={d\sigma_{\rm pAu}}/{dydM^2}$. If $\cos\left(X,Y\right) \sim (-)1$, the two quantities X and Y are strongly (anti)correlated whereas if $\cos\left(X,Y\right) \sim 0$, the two are independent. In computing the correlation cosine, we have kept the proton PDF $f_i^{\rm p}$ fixed to the CT14NLO central set, and used the CT14NLO error sets to vary $f_i^{\rm Au}$. In other words, we compute the cross sections using $f_i^{p}=f_i^{p}(S_0)$ for the proton and $f_i^{\rm p/Au}(S_k^\pm) = f_i^{\rm p}(S_k^\pm) R_i^{\rm Au, EPPS16}{(S_0)}$ to form Eq.~(\ref{eq:correl}). The point in using the CT14NLO error sets is that the CT14NLO fit function is somewhat more flexible at small $x$ than the EPPS16 ansatz, so this should give a better estimate of the true correlations. From Fig.~\ref{fig:correlationcosine} we see that the gluon distribution at small $x$ is anticorrelated with the forward Drell-Yan cross sections and at larger $x$ we see a positive correlation. The main reason for the small-$x$ anticorrelation is the direct contribution from the quark-gluon scattering, present at NLO and beyond, which is negative and clearly non negligible. In our case, this amounts to $\sim 15\ldots 45\%$ of the cross sections with all partonic channels included. This contribution becomes increasingly important towards low $x_2$ and higher $M$. The large-$x$ positive correlation persists also in a leading-order calculation so it is due to the indirect constraints from the scale evolution and momentum sum rule. 
Because the $q\overline{q}$ channel dominates the cross sections, the correlation with gluon PDF is moderate but can reach almost up to $\sim$40\% at small $x$. Below $x \sim 10^{-3}$ the correlation begins to decrease as this region is beyond the kinematic reach of the projected experimental acceptance. In part, the residual non-zero correlations $x \lesssim 10^{-3}$ are due to the assummed form of the small-$x$ fit function, but the momentum conservation and evolution effects also place indirect constraints. All in all, we can conclude that the Drell-Yan production at forward kinematics is indeed sensitive to the small-$x$ gluon PDFs.

We note that the dijet and photon-jet pseudodata probe the mid- and high-$x$ part of the nuclear PDFs. The uncertainties for these two observables are dominated by the assumed 5\% uncorrelated bin-to-bin systematic error and the obtained improvement in nuclear PDFs are dictated by this assumption. If systematic uncertainties like those achieved in p+Pb collisions at the LHC \cite{Sirunyan:2018qel} could be reached, the impact would be clearly larger. In addition, the systematic uncertainty of the LHC measurements is almost always of correlated nature, but such correlation is difficult to estimate in advance. All in all, assuming a 5\% uncorrelated systematic uncertainty appears thus a reasonable test scenario which should not overstate the constraining power.

\section{Constraining the $A$ Dependence of nuclear PDFs with Lighter Ions}

\begin{figure}[htb!]
    \centering
    \includegraphics[width=0.80\linewidth]{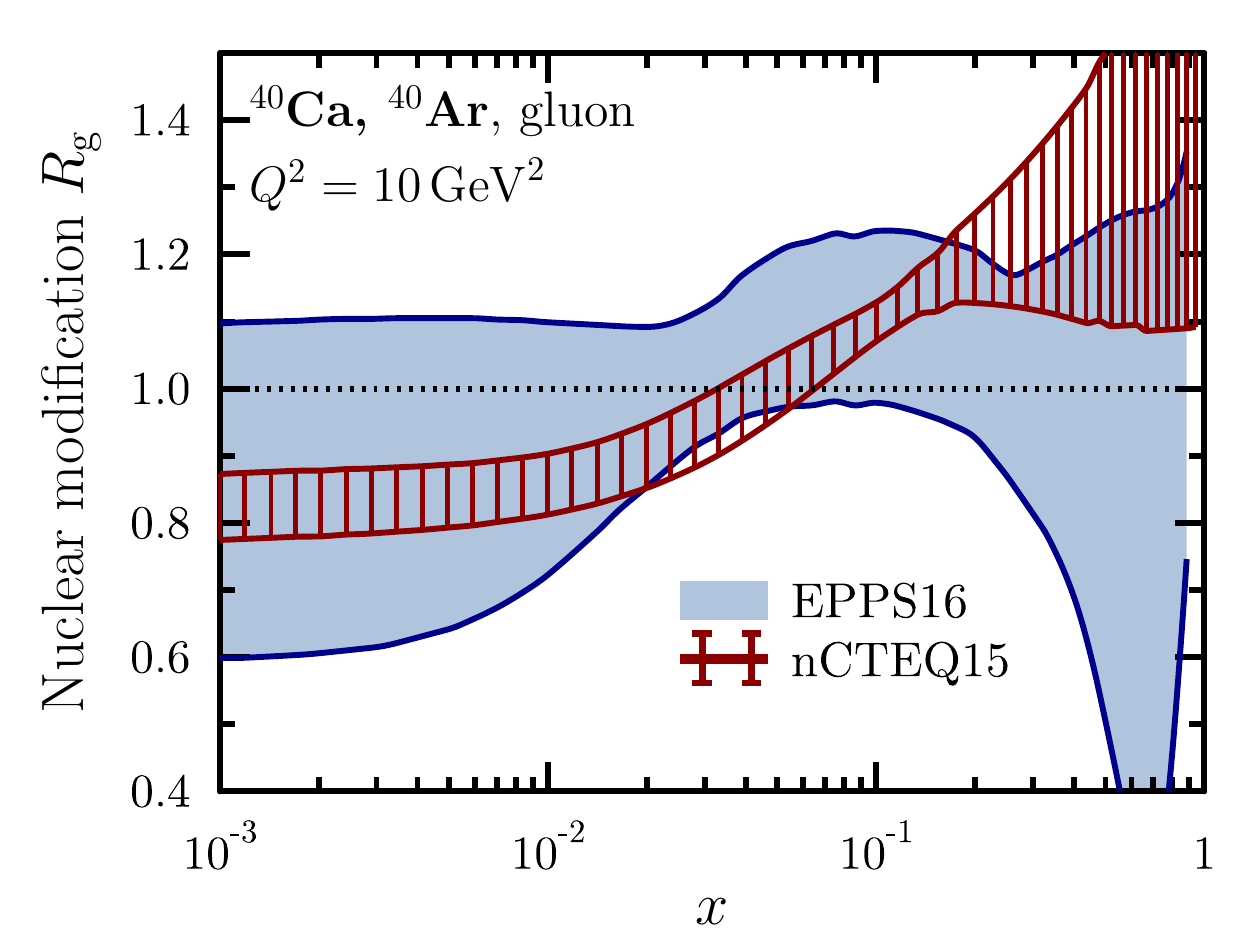}    \caption{Nuclear modifications of gluon PDF for $A=40$ nuclei from EPPS16 and nCTEQ15.}
    \label{fig:Ca_glue}
\end{figure}

\begin{figure}[htb!]
    \centering
    \includegraphics[width=0.75\linewidth]{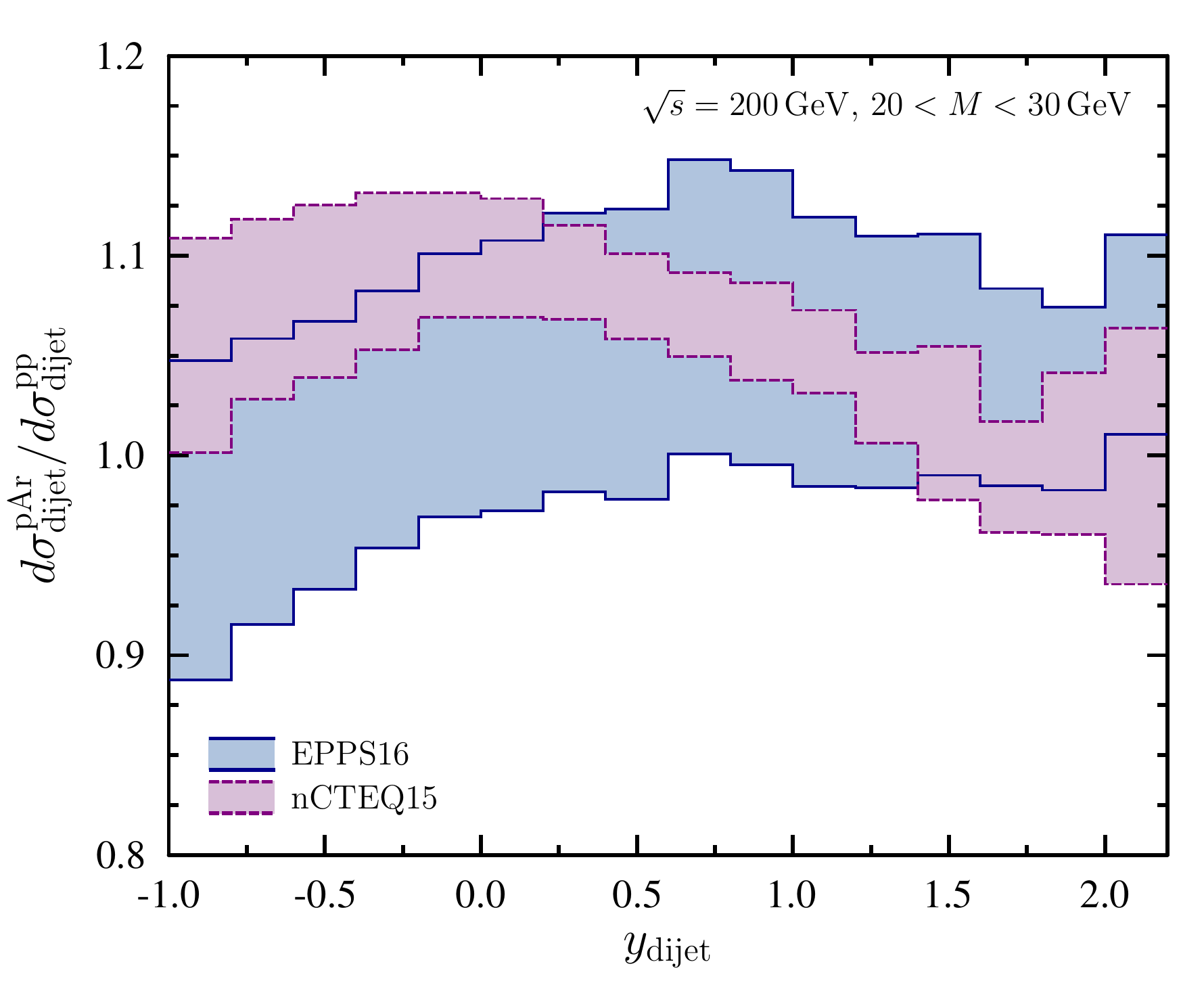}  
    \caption{Dijet nuclear modification in p-Ar scattering as predicted by EPPS16 and nCTEQ15 for invariant mass $20<M<30$.}
    \label{fig:dijetAr}
\end{figure}

The mass-number ($A$) dependence of the current nuclear PDFs is not well known --- direct constraints exist only for large-$x$ valence quarks and intermediate-$x$ sea-quarks. On one hand, e.g. in the EPPS16 analysis, the guideline has been that the nuclear effect should be larger for larger nuclei at the parametrization scale $Q=m_{\rm charm}$ which then tends to lead to physically sound $A$ systematics also at larger $Q$. On the other hand, in the recent nuclear-PDF analysis by the NNPDF collaboration \cite{AbdulKhalek:2019mzd} there is less direct control over the $A$ dependence and thus the nuclear effects from one nucleus to another can fluctuate significantly. Due to the p+Pb and Pb+Pb collisions program at the LHC, the near-future improvements on nuclear PDFs are bound to be driven by the Pb nucleus. For example, the dijet \cite{Sirunyan:2018qel}, D-meson \cite{Aaij:2017gcy} and W$^\pm$ \cite{CMS:2018ilq} measurements efficiently constrain \cite{Kusina:2017gkz, Eskola:2019dui} the gluons in the Pb nucleus, perhaps providing even stronger constraints for Pb than what we have found in the present study for Au. The LHCb fixed-target mode facilitates measurements on lighter noble-gas targets \cite{Aaij:2018ogq}, but only the very-high $x$ regime of nuclear PDFs is accessible. However, e.g. in astrophysical applications the relevant nuclei (e.g. oxygen and nitrogen) are much lighter and thus collider measurements involving lighter nuclei would be very much useful \cite{Aaij:2018svt}. In addition, the study of the onset of jet quenching and saturation phenomena with lighter ions will require nuclear PDFs for nuclei other than Au or Pb. Interests towards light-ion beams at the LHC have been expressed \cite{Citron:2018lsq} but since the main focus of LHC is still in p+p collisions, it is not clear whether and when this would materialize. Here, the flexibility of RHIC to run with different ions is a clear asset. Indeed, at least p, d, Al, Cu, Ru, Zr and U ions have already been used in physics runs which demonstrates that a proper \emph{``$A$ scan''} is, in principle, possible. The same multi-ion option would also be available if RHIC is eventually turned into an Electron-Ion Collider \cite{Accardi:2012qut}, where the possibilities to constrain nuclear PDFs are undisputed \cite{Aschenauer:2017oxs,Aschenauer:2017jsk}. To highlight the present uncertainties for light ions, Fig.~\ref{fig:Ca_glue} shows the nuclear effects for $A=40$ (Ca, Ar) from the EPPS16 and nCTEQ15 \cite{Kovarik:2015cma} global fits of nuclear PDFs. While the uncertainty bands overlap, the shape at intermediate and large $x$ are quite different; while the nuclear effects in nCTEQ15 monotonically rise towards high $x$, the EPPS16 error band more closely resembles the typical pattern of shadowing, antishadowing and EMC-effect. Fig.~\ref{fig:dijetAr} shows how this different behaviour would be reflected in dijet production. In the backward direction ($y_{\rm dijet} < 0$) one is sensitive to the large-$x$ part of nuclear PDFs and the nCTEQ15 prediction tends to be above the EPPS16 one, consistently with Fig.~\ref{fig:Ca_glue}. The difference in Fig.~\ref{fig:dijetAr} is not as marked as in Fig.~\ref{fig:Ca_glue} as towards large $x$ the valence quarks also play an increasingly important role. Towards $y_{\rm dijet} \gg 0$ the probed $x$ gets lower and, in line with Fig. ~\ref{fig:Ca_glue}, the nCTEQ15 prediction tends to be at the lower limit of EPPS16. Assuming a similar $\sim$50\% reduction in the gluon PDF uncertainties as found for Au in Fig. ~\ref{fig:fit_nPDF}, it appears reasonable that the measurements would be able to resolve between nCTEQ15 and EPPS16. In an approach like that of the NNPDF collaboration \cite{AbdulKhalek:2019mzd}, where more freedom for the $A$ dependence is allowed than in nCTEQ15 or EPPS16, the benefit would be even more pronounced.

An additional interesting possibility we would like to point out would be to study p+$A_i$ collisions of two isobaric nuclei $A_1$ and $A_2$ (e.g. $A_1=^{96}_{44}$Ru vs. $A_2=^{96}_{40}$Zr collisions) with constant $A$ but differences in proton and neutron content. Precision measurements of e.g. (p+Ru)/(p+Zr) ratios for hard processes (like those discussed in this paper) would allow a study of the assumptions made in the present global fits of nuclear PDFs. Indeed, it is currently assumed that the nuclear effects depend only on the mass number $A$, and not on the mutual balance of neutrons and protons. In addition, the isospin symmetry (i.e. $u^{\mathrm{proton}/A} = d^{\mathrm{neutron}/A}$) is assumed to be exact. Thus, (p+Ru)/(p+Zr) ratios, or other similar constant-$A$ combinations, would test the assumptions made in global analyses at a deeper level and also test other theoretical approaches, e.g. the importance of short-range nucleon-nucleon correlations \cite{Hen:2012fm}, or the lack of them \cite{Arrington:2019wky}. In principle, in an optimal situation the neutron-to-proton mixture in the two nuclei should be as different as possible, with (at least nearly) constant $A$. Such a measurement makes optimal use of the flexibility of the RHIC facility. 

\section{Summary}

Using the Au nucleus as a test case, we have examined the prospects for constraining nuclear gluon PDFs at RHIC with new measurements that assume detector acceptances similar to those proposed for STAR and sPHENIX with forward upgrades. We have found that the Drell-Yan process at low invariant mass is able to significantly constrain the low $x$ gluon distribution with up to 50\% reduction in the current EPPS16 uncertainty. The constraints at higher $x$ depend considerably on the assumed systematic uncertainty, which is expected to dominate over the statistical uncertainty for dijet and photon-jet processes. Assuming an order of 5\% bin-to-bin independent systematic uncertainty leads to modest constraints at mid- and high-$x$ region. Even so, we find the inclusion of additional observables along with the Drell-Yan data of utmost importance to overcome the overall normalization uncertainty in the $R_{\mathrm{p}A}$ ratio. Without supplementing the Drell-Yan pseudodata with other observables (here either dijets, photon-jet, or both), we find that the power of the measurement of Drell-Yan  to constrain the small-$x$ behavior of the gluon is lost.  
It is possible that even stronger constraints could be obtained if measurements of forward direct photons could be added to this suite of observables. 

While the focus of our analysis was on the Au nucleus, similar constraints can be expected to be obtained for any other nucleus. In this respect we briefly discussed the $A$ dependence of nuclear PDFs and highlight the significant opportunity for improvements that could be attained with a proper \emph{$A$-scan} -- measuring the same observables with several nuclear beams -- for which the RHIC collider provides a unique opportunity.

\begin{acknowledgments}
I.~H. and H.~P. acknowledge the support from the Academy of Finland, Project 308301. The work of P.~P. has been supported by the Magnus Ehrnrooth Foundation. The Finnish IT Center for Science (CSC) is acknowledged for super-computing resources under the Project jyy2580. J.~O. acknowledges support from the Margaret and Herman Sokol Faculty Awards of University of Michigan and the Office of Nuclear Physics in the Office of Science of the Department of Energy under Grant No. DE-SC0013393. J.~L. acknowledges support from the Office of Nuclear Physics in the Office of Science of the Department of Energy under Grants No. DE-FG02-10ER41719 and DE-FG02-92ER40962. 
\end{acknowledgments}

\bibliography{cites}

\end{document}